\documentclass[aps,prd,nofootinbib, groupedaddress,superscriptaddress, twocolumn]{revtex4-2}

\pdfoutput=1
\usepackage{rotating}
\usepackage{array,makecell,multirow}

\usepackage[T1]{fontenc}
\usepackage{hyperref}
\usepackage{xcolor}
\usepackage{graphicx}
\usepackage{array,booktabs}
\usepackage{float}
\usepackage{amsmath}
\usepackage{amssymb}
\usepackage{multirow} 
\usepackage{lineno}
\usepackage{tabularx}
\usepackage{comment}
\usepackage[normalem]{ulem}
\usepackage[caption=false]{subfig}

\newcommand{\APC}{APC, Universit\'e de Paris, CNRS, Astroparticule et Cosmologie, Paris F-75013, France}
\newcommand{\AQLNGS}{INFN Laboratori Nazionali del Gran Sasso, Assergi (AQ) 67100, Italy}
\newcommand{\AQGSSI}{Gran Sasso Science Institute, L'Aquila 67100, Italy}

\newcommand{\AstroCeNT}{AstroCeNT, Nicolaus Copernicus Astronomical Center, 00-614 Warsaw, Poland}
\newcommand{\Augustana}{Physics Department, Augustana University, Sioux Falls, SD 57197, USA}
\newcommand{\Belgorod}{Radiation Physics Laboratory, Belgorod National Research University, Belgorod 308007, Russia}
\newcommand{\BHSU}{School of Natural Sciences, Black Hills State University, Spearfish, SD 57799, USA}

\newcommand{\CAUniPHY}{Physics Department, Universit\`a degli Studi di Cagliari, Cagliari 09042, Italy}
\newcommand{\CAINFN}{INFN Cagliari, Cagliari 09042, Italy}

\newcommand{\CPPM}{Centre de Physique des Particules de Marseille, Aix Marseille Univ, CNRS/IN2P3, CPPM, Marseille, France}
\newcommand{\CTLNS}{INFN Laboratori Nazionali del Sud, Catania 95123, Italy}
\newcommand{\ENUniCEE}{Engineering and Architecture Faculty, Universit\`a di Enna Kore, Enna 94100, Italy}

\newcommand{\FNAL}{Fermi National Accelerator Laboratory, Batavia, IL 60510, USA}

\newcommand{\GEUni}{Physics Department, Universit\`a degli Studi di Genova, Genova 16146, Italy}
\newcommand{\GEINFN}{INFN Genova, Genova 16146, Italy}

\newcommand{\Hawaii}{Department of Physics and Astronomy, University of Hawai'i, Honolulu, HI 96822, USA}
\newcommand{\Houston}{Department of Physics, University of Houston, Houston, TX 77204, USA}
\newcommand{\IHEP}{Institute of High Energy Physics, Beijing 100049, China}

\newcommand{\JINR}{Joint Institute for Nuclear Research, Dubna 141980, Russia}
\newcommand{\Krakow}{M. Smoluchowski Institute of Physics, Jagiellonian University, 30-348 Krakow, Poland}
\newcommand{\Kurchatov}{National Research Centre Kurchatov Institute, Moscow 123182, Russia}

\newcommand{\LNFINFN}{INFN Laboratori Nazionali di Frascati, Frascati 00044, Italy}
\newcommand{\LNHB}{Universit\'e Paris-Saclay, CEA, List, Laboratoire National Henri Becquerel (LNE-LNHB), F-91120 Palaiseau, France}

\newcommand{\LPNHE}{LPNHE, CNRS/IN2P3, Sorbonne Universit\'e, Universit\'e Paris Diderot, Paris 75252, France}
\newcommand{\Manchester}{The University of Manchester, Manchester M13 9PL, United Kingdom}
\newcommand{\MEPhI}{National Research Nuclear University MEPhI, Moscow 115409, Russia}

\newcommand{\MIINFN}{INFN Milano, Milano 20133, Italy}

\newcommand{\MIUni}{Physics Department, Universit\`a degli Studi di Milano, Milano 20133, Italy}
\newcommand{\MSU}{Skobeltsyn Institute of Nuclear Physics, Lomonosov Moscow State University, Moscow 119234, Russia}
\newcommand{\NAINFN}{INFN Napoli, Napoli 80126, Italy}
\newcommand{\NAUniPHY}{Physics Department, Universit\`a degli Studi ``Federico II'' di Napoli, Napoli 80126, Italy}

\newcommand{\Petersburg}{Saint Petersburg Nuclear Physics Institute, Gatchina 188350, Russia}
\newcommand{\PGUniCBB}{Chemistry, Biology and Biotechnology Department, Universit\`a degli Studi di Perugia, Perugia 06123, Italy}
\newcommand{\PGINFN}{INFN Perugia, Perugia 06123, Italy}
\newcommand{\PIINFN}{INFN Pisa, Pisa 56127, Italy}
\newcommand{\PIUniPHY}{Physics Department, Universit\`a degli Studi di Pisa, Pisa 56127, Italy}
\newcommand{\PNNL}{Pacific Northwest National Laboratory, Richland, WA 99352, USA}
\newcommand{\Princeton}{Physics Department, Princeton University, Princeton, NJ 08544, USA}

\newcommand{\RHUL}{Department of Physics, Royal Holloway University of London, Egham TW20 0EX, UK}
\newcommand{\RMTreINFN}{INFN Roma Tre, Roma 00146, Italy}
\newcommand{\RMTreUni}{Mathematics and Physics Department, Universit\`a degli Studi Roma Tre, Roma 00146, Italy}
\newcommand{\RMUnoINFN}{INFN Sezione di Roma, Roma 00185, Italy}
\newcommand{\RMUnoUni}{Physics Department, Sapienza Universit\`a di Roma, Roma 00185, Italy}

\newcommand{\SSUniCHP}{Chemistry and Pharmacy Department, Universit\`a degli Studi di Sassari, Sassari 07100, Italy}

\newcommand{\UCDavis}{Department of Physics, University of California, Davis, CA 95616, USA}
\newcommand{\UCLA}{Physics and Astronomy Department, University of California, Los Angeles, CA 90095, USA}
\newcommand{\UMass}{Amherst Center for Fundamental Interactions and Physics Department, University of Massachusetts, Amherst, MA 01003, USA}

\newcommand{\USP}{Instituto de F\'isica, Universidade de S\~ao Paulo, S\~ao Paulo 05508-090, Brazil}
\newcommand{\VTech}{Virginia Tech, Blacksburg, VA 24061, USA}

\begin{document}

\title{Search for low-mass dark matter WIMPs with 12 ton-day exposure of DarkSide-50}

\author{P.~Agnes}\affiliation{\RHUL}
\author{I.F.M.~Albuquerque}\affiliation{\USP}
\author{T.~Alexander}\affiliation{\PNNL}
\author{A.K.~Alton}\affiliation{\Augustana}
\author{M.~Ave}\affiliation{\USP}
\author{H.O.~Back}\affiliation{\PNNL}
\author{G.~Batignani}\affiliation{\PIINFN}\affiliation{\PIUniPHY}
\author{K.~Biery}\affiliation{\FNAL}
\author{V.~Bocci}\affiliation{\RMUnoINFN}
%\author{G.~Bonfini}\affiliation{\AQLNGS}
\author{W.M.~Bonivento}\affiliation{\CAINFN}
% ...

\collaboration{The DarkSide Collaboration}\noaffiliation

%\date{\today}

\begin{abstract}
% 0.64~keV$_{nr}$ or

We report on the search for dark matter WIMPs in the mass range below 10~GeV/c$^2$, from the analysis of the entire dataset acquired with a low-radioactivity argon target by the DarkSide-50 experiment at LNGS.  The new analysis benefits from more accurate calibration of the detector response, improved background model, and better determination of systematic uncertainties, allowing us to accurately model the background rate and spectra down to  0.06~keV$_{er}$. A 90\% C.L. exclusion limit for the spin-independent cross section of   3~GeV/c$^2$ mass WIMP on nucleons is set at 6$\times$10$^{-43}$~cm$^2$, about a factor 10 better than  the previous DarkSide-50 limit. 
This analysis extends the exclusion region  for spin-independent dark matter interactions below the current experimental constraints   in the  $[1.2, 3.6]$~GeV/c$^2$ WIMP mass range.
\end{abstract}

\maketitle

%introduction
\section{Introduction}
In the last decade, the noble-liquid dual-phase time projection chamber (TPC)  has emerged as leading detection technology  in the search for Weakly Interacting Massive Particles (WIMPs) \cite{LUX:2016ggv,DarkSide:2018kuk, XENON:2018voc, PandaX-4T:2021bab},  among the best-motivated  dark matter candidates \cite{Lee:1977ua}, with  mass  above 10~GeV/c$^2$. The strengths of this approach are the intrinsic radiopurity and  scalability of the target, and the accurate topological reconstruction of interacting particles by  detection of both scintillation and ionization signals.

Such detectors also exhibit world class sensitivity in the search for light dark matter candidates, such as GeV/c$^2$ mass scale WIMPs \cite{DarkSide:2018bpj,  DarkSide:2018ppu, LUX:2018akb, XENON:2019zpr, XENON:2019gfn} and Axion-Like Particles (ALPs) \cite{XENON100:2014csq, LUX:2017glr, PandaX:2017ock} when exploiting the ionization signal alone to reach detection thresholds in the keV range. Although the scintillation signal is no longer observable in this regime, dual-phase TPCs drift and extract single ionization electrons in  gas  with near 100\% efficiency~\cite{Bondar:2009rd}. Electron signal is then amplified in gas by a factor up to $\sim$20~\cite{DarkSide:2018bpj}  exploiting electroluminescence,  generated in the transit of charged particles in the gas phase under a strong electric field. This amplification guarantees the possibility to trigger on single electron signals  and centimeter level resolution in the reconstruction of the interaction position on the plane orthogonal to the electric field. Finally,  multi-site signals corresponding to multiple scattering particles, not compatible with those induced by dark matter, are efficiently rejected \cite{XENON:2016jmt, LUX:2018akb, DarkSide:2018bpj}. 

Liquid argon (LAr)  and xenon  detectors have fairly similar performance in terms of ionization yield. However, because of the lower atomic mass, low-mass WIMP scattering off argon produces more energetic recoils, with a higher probability of being detected above the threshold. This compensates for the lower cross section in argon compared to xenon and makes DarkSide-50     the most  sensitive experiment to date to WIMP interactions in the  [1.8,  3.0] GeV/c$^2$ mass range, with a fiducial LAr target of  only $\sim$20 kg \cite{DarkSide:2018bpj}. 

\begin{figure}[h]
\includegraphics[width=1.0\linewidth]{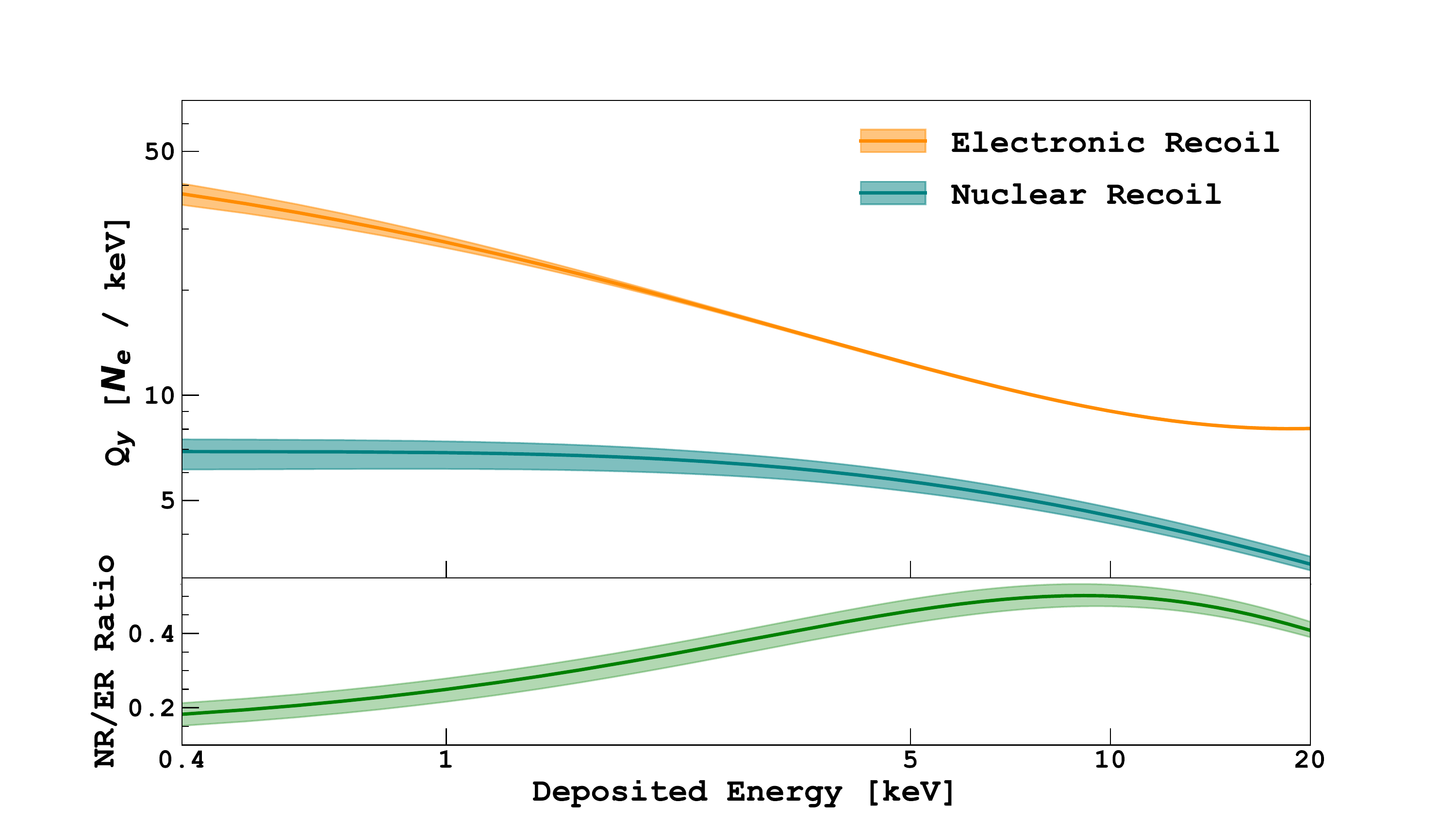}
\caption{Comparison and ratio between  LAr ionization responses to nuclear (NR) and electronic (ER) recoils, as a function of the deposited energy. The two energy scales were measured using DarkSide-50 data and datasets from the ARIS~\cite{Agnes:2018mvl} and SCENE~\cite{SCENE:2014iyj} experiments, and are reported in ref.~\cite{DarkSide:2021bnz}.}
\label{fig:qy}
\end{figure}

Recently, the DarkSide-50 Collaboration  re-analyzed data from calibration campaigns with radioactive sources and measured, with high-accuracy, the LAr ionization response (Q$_{y}$), shown in Fig.~\ref{fig:qy},  to  electron (ER) and nuclear (NR) recoils  down to a few hundred electron-volts \cite{DarkSide:2021bnz}. This measurement represents the first step of a comprehensive re-analysis of the entire DarkSide-50 data-set to provide improved constraints on low-mass WIMP-nucleon interactions.  In this work, we present an improved data selection that, together with a more accurate background model, greatly impact the experimental constraints on low-mass WIMP from the DarkSide-50 experiment. Other elements of the re-analysis are an improved detector response model and a refined treatment of systematics into the statistical analysis,  discussed in detail in the next sections.

\section{Detector}
The DarkSide-50 experiment operated between 2013 and 2019  in the Hall C of the Gran Sasso National Laboratory (LNGS) in Italy.  The first data-taking campaign ran from November 2013 to April 2015 with an atmospheric argon (AAr) target, then replaced with low-radioactivity argon extracted from deep underground  (from now on named underground argon),  with reduced  activity of cosmogenic isotopes \cite{DarkSide:2015cqb}. The TPC active mass is  46.4$\pm$0.7~kg. The uncertainty on the active mass is primarily due to the thermal contraction of the polytetrafluoroethylene (PTFE), which defines the TPC cylindrical volume. The PTFE sidewalls are  surrounded by field shaping copper rings. These provide a uniform 200~V/cm electric field in the liquid bulk. Two arrays of 19 3'' diameter photomultiplier tubes (PMTs),  installed at both ends of the TPC behind  transparent anode and cathode windows, observe light from  scintillation in the liquid (S1) and from the electroluminescence in the gas  (S2). Electroluminescence occurs when ionization electrons, after being drifted across the liquid bulk, are extracted with a 2.8 kV/cm electric field in the gas phase,  and here drifted thanks to a 4.2 kV/cm electric field. All the inner surfaces of the TPC   are coated with tetraphenyl butadiene (TPB), a wavelength shifter that absorbs 128~nm  photons from argon de-excitation and re-emits photons, whose  wavelengths are peaked at 420~nm.  More details about the TPC are reported in ref.  \cite{DarkSide:2014llq, DarkSide:2015cqb}.  

The TPC is hosted inside a 120~l double-wall cryostat and is shielded by a liquid scintillator and a water Cherenkov detector against neutrons and cosmic muons, respectively. All three detectors are read out upon a trigger from the TPC that requires at least two PMTs above a threshold of 0.6 photoelectrons  in coincidence within 100~ns \cite{DarkSide:2017odo}.

\section{Data selection}
The dataset reported in this paper consists of 653.1~live-days of underground argon  data, taken from December~12,~2015, to February~24,~2018, with an average trigger rate of 1.54~Hz. 

WIMPs may eventually scatter only once in the LAr target because of the extremely low cross section. Since each particle interaction is associated with an S2 pulse, only events with a single S2 pulse are considered for this analysis.   Given the low-energy regime and the low detection efficiency of S1 photons (0.16$\pm$0.01 \cite{DarkSide:2017wdu}), not all of these  events have an associated S1 pulse. Therefore,  selected events are divided into two categories, depending on whether they have one (S2-only) or two pulses (S1 and S2).  The only exception  is made in the presence of ``echoes'', i.e. electrons  extracted by S1 or S2 128~nm photons from the cathode, via  photoelectric effect~\cite{DarkSide-50:2021wku}. Events with echoes are  efficiently identified by looking at the time coincidence between the echo and the pulse that induced it, equals to the maximum drift time (376~$\mu$s) \cite{DarkSide-50:2021wku}.

The DarkSide-50 position reconstruction algorithm on the plane orthogonal to the electric field is inefficient at the keV scale, the region of interest for this analysis. For this reason, the event position is here defined as the position of the top-array PMT observing the largest fraction of S2 photons. Based on this definition, events selected by the outermost ring of PMTs are discarded as they fall in the volume most exposed to external radioactive contamination. %from the cryostat.
The signal acceptance\footnote{From now on, the term acceptance will be referred to as signal acceptance.} of this cut corresponds to 41.9\% of the entire volume, and was probed to be independent on the size of the S2 pulse with Monte Carlo simulations~\cite{DarkSide:2017wdu}. More details on this volume fiducialization can be found in ref.~\cite{DarkSide:2021bnz}. 

The measured S2 yield, defined as the mean number of photoelectrons per ionization electron extracted in the gas pocket, is  23$\pm$1~pe/e$^-$, for events localized beneath the central PMT \cite{DarkSide:2018bpj}. The radial dependence of the S2 yield, already discussed in ref.~\cite{DarkSide:2017wdu}, is here corrected to the value at the center of the TPC, using a correction map extracted from $^{83m}$Kr calibration data~\cite{DarkSide:2021bnz}. The energy observable used in this analysis is the number of detected electrons, $N_e$, defined as the corrected number of S2 photoelectrons  divided by the S2 yield\footnote{In case of ERs, the number of electrons is the sum of primary  and ionization electrons.}. The energy range of interest for this analysis is defined from 4  to 170 $N_e$, corresponding to [0.06, 21]~keV$_{er}$ ([0.6, 288]~keV$_{nr}$) in the ER (NR) energy scale. The upper limit is defined up to where the energy scale calibrations have been validated.  The trigger efficiency is estimated at 100\% in the full range of interest \cite{DarkSide:2018bpj}. The lower bound of the region of interest is chosen in order to avoid contamination from spurious ionization electrons trapped by trace impurities and then released, as discussed later.

The data selection relies on two classes of  cuts: quality cuts, defined to  reject pulse pile-ups, and selection cuts, to remove   spurious electrons,  alpha-induced events, and events with an anomalous start time. Cut efficiencies and acceptances are estimated either via Monte Carlo   or on the  AAr sample. The latter is  dominated by $^{39}$Ar, whose activity is three orders of magnitude higher than the underground argon (UAr) campaign event rate. AAr is an optimal calibration sample since $^{39}$Ar $\beta$-decays are detected as  single-sited interactions, like the  signature expected from dark matter interactions.

\subsection{Quality cuts}

S1 and S2 pulses are identified with a fixed threshold on the overall waveform obtained by summing all individual waveform  as described in ref.~\cite{DarkSide:2014llq}. S2 pulses are  distinguished  from S1 ones by requiring  $f_{90}$, the fraction of light observed in the first 90~ns, to be less than 0.1.  This cut has been checked on Monte Carlo simulation to be full  efficiency already for signals equivalent to 4~$N_e$.

The S2 sample selected by $f_{90}$ $<$0.1 may however contain  a fraction of events where the signal is actually composed by several pulses overlapping in time.  A series of ``quality'' cuts, based on the time profile and topological distribution of S2 photons, were then implemented to reject such contamination.

The first quality cut requires that the identified S2 pulse is contained in  100~$\mu$s.  Signals longer than  100~$\mu$s  are associated to events with overlapping pulses, which are not resolved by the pulse finder algorithm. 

Other events with unresolved overlapping pulses are rejected by looking at the S2 pulse peak time, the time when the pulse reaches its maximum amplitude relative to the pulse start time. Electron diffusion in LAr along the drift increases the longitudinal size of the drifting ionization cloud. This induces longer S2 signals for events near the cathode, with a higher probability that two distinct but nearby energy deposits are merged in the reconstruction. When this happens, the superposition of signals produces an S2 pulse with a peak time that is delayed compared with single scattering events. Requiring the S2 peak time to be less than 6~$\mu$s reduces this background component.

Pile-up of S1 and S2 pulses are due to the pulse finder algorithm being unable to separate two pulses closer than 2~$\mu$s. This condition occurs  approximately within 2 mm from the grid, about 0.5\% of the entire volume. These events are rejected if the peak time of the S2 pulse is less than 200~ns or if the FWHM 
 is less than 100~ns. The  acceptance is estimated at 99.75$\pm$0.25\%.

In the LAr  volume below 2~mm from the grid, the combined acceptance of the S2 peak time, S2 FWHM, and S2 gate  cuts  is estimated via Monte Carlo at $\sim$95$\%$ at 4~$N_e$ and  $\sim$100$\%$ at equal to or larger than 15~$N_e$.

\begin{figure}[h]
\includegraphics[width=1.0\linewidth]{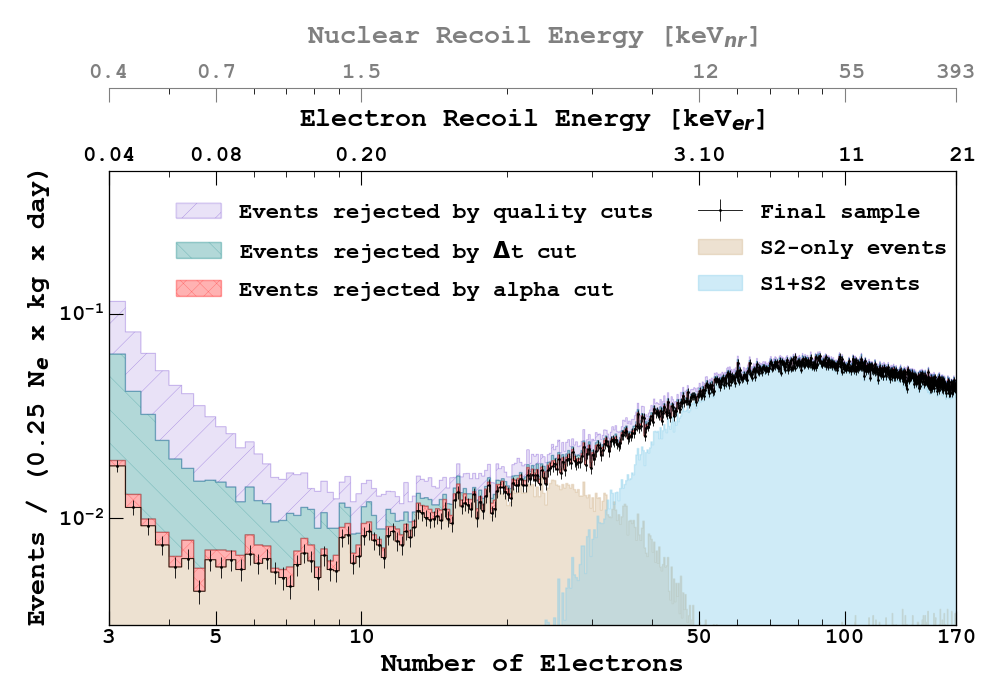}
\caption{$N_e$ spectra at different steps of the data selection,  after rejection of events outside the fiducial volume and with multiple interactions.}
\label{fig:data_selection}
\end{figure}

The impact of the quality cuts on the ionization electron spectrum  is shown in Fig.~\ref{fig:data_selection}. 

Some of the  S2-only pulses are observed with an anomalous pulse start time, preceding the trigger time  by several microseconds. The probability that the photons from the tail of a S2 pulse   trigger the detector is estimated  by means of electronics simulation~\cite{DarkSide:2017wdu} and event reconstruction and decreases as $N_e$ decreases. At 3~e$^-$, at the low threshold of the region of interest, the probability that S2 pulses trigger the detector after 1~$\mu$s (1.8~$\mu$s) from the start time of the pulse itself is less than 1\% (0.1\%), while we observe events with  start time by up to  7~$\mu$s,  preceding the trigger time.   A selection on S2 pulse start time was designed  on the simulated sample to reject such events from the S2-only sample, while keeping 99\% acceptance constant as a function of $N_e$.

\subsection{Selection cuts}

The sample resulting from the quality cuts is contaminated by two classes of background events, appropriately rejected by the optimized selection cuts. The first category corresponds to spurious ionization electrons trapped along their drift by trace impurities or at the liquid surface and released with delays of up to hundreds of milliseconds. These produce signals equivalent to up to a few electrons, as discussed in ref.~\cite{DarkSide:2018bpj}. 

%A cut is implemented to suppress such contribution  by exploiting the time correlation with the preceding event . by vetoing  

%A cut is implemented to suppress such contribution by vetoing  events that triggered the DAQ less than 20~ms from the previous one. 
 A veto of 20~ms from after each event triggering the DAQ is implemented to suppress such contribution, exploiting the time correlation with the preceding event. This cut extends the one of 2~ms used in the 2018 analysis~\cite{DarkSide:2018bpj},  introduces a deadtime of 3\%, evaluated from the rate of uncorrelated events, fitted from the distribution of the event time differences.

\begin{figure}[h]
\begin{center}
\includegraphics[width=1.0\linewidth]{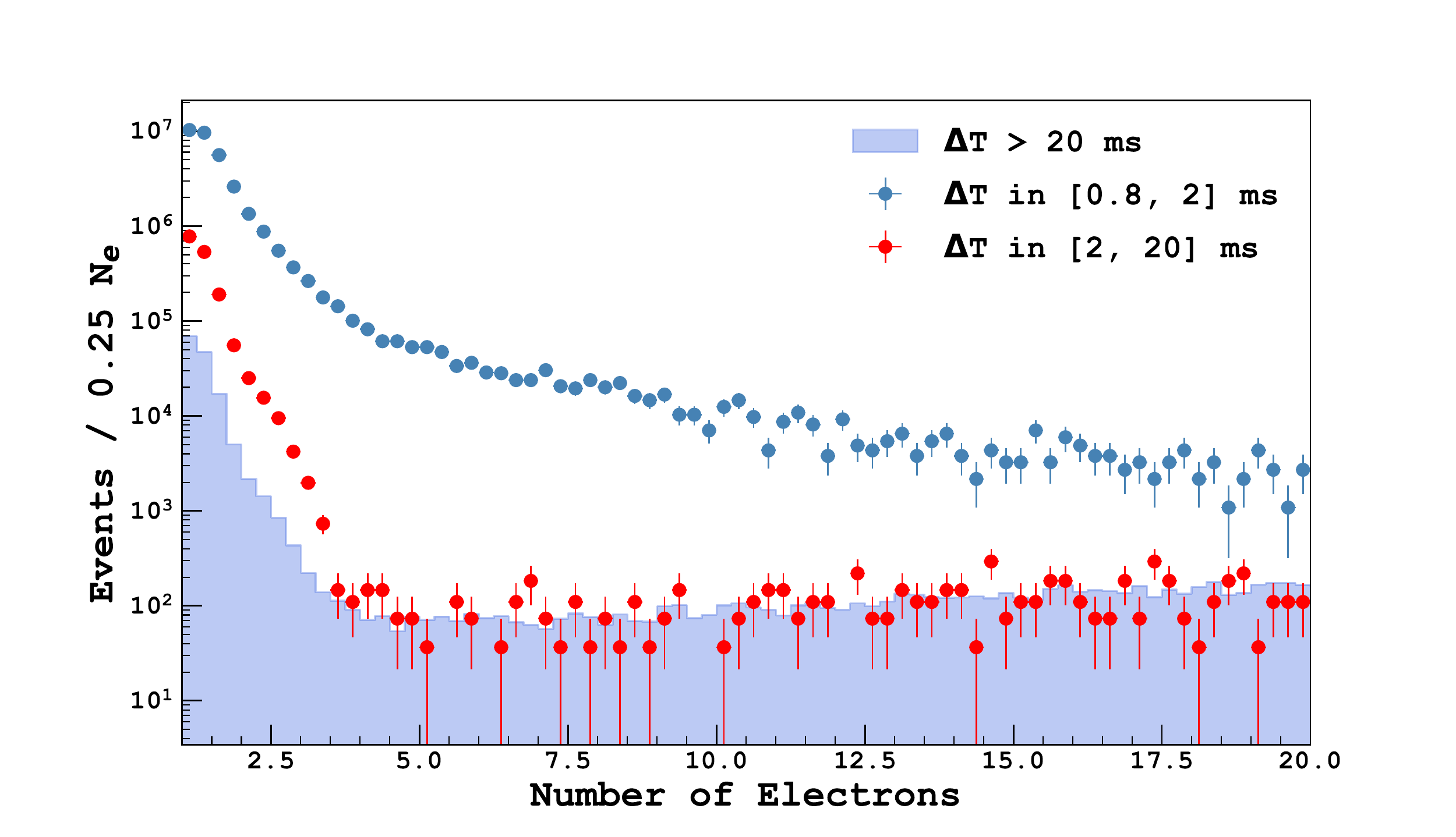}
\caption{$N_e$ spectra after selection cuts  requiring a time coincidence ($\Delta$T) with the preceding event higher than 20~ms (shaded blue), between [0.8, 2]~ms (blue dots) and between [2, 20]~ms (red dots). The  spectra are normalized to livetime of the sample with $\Delta$T$>$20~ms, whose spectrum is statistically compatible  with the one with $\Delta$T in [2, 20]~ms above 4~$N_e$, where the contamination from spurious correlated electrons becomes negligible. }
\label{fig:low_ne}
\end{center}
\end{figure}

Figure~\ref{fig:low_ne} shows a comparison of spectra of the event rejected by the veto cut and the remaining events normalized to livetime. A significant suppression of low $N_e$ events is obtained after veto. Remaining background is from spurious electron with longer time delay ($\Delta$T) and from a component of this background without a clear time correlation with previous events. The $N_e$ spectra for samples selected  with $\Delta$T in [2, 20]~ms and $>$20~ms  are in agreement above 4~$N_e$, indicating that the contribution of  spurious electrons  is negligible in the analysis range. In addition, the probability of a single-scatter event being misidentified as multiple-scatters one due to the fast re-emission of a spurious electron was estimated to be negligible.  

%indicating that the expected contribution of the spurious electron background is negligible in the analysis range. 

A second class of backgrounds is associated to  events  characterized by a large S1, paired with an anomalously low S2 pulse. Their  origin has been traced to $\alpha$ particles generated inside the TPC walls,  at shallow depth near the interface with the LAr bulk. The induced ionization electrons are   absorbed by the walls themselves but scintillation photons may extract additional electrons from the cathode by  photoelectric effect. To reject these events, we developed a specific cut affecting the S1+S2 sub-sample only, tuned on calibration data.   The cut is built by fitting with a normal distribution log$_{10}$(S2/S1), for each slice of log$_{10}$(S1), and removing events below and above 2.58$\sigma$ from the mean, corresponding to 99\% acceptance. 

The overall acceptance from both the quality and the selection cuts,   almost flat with respect to the recoil energy,  varies from 38\% at 4~e$^-$ to 40\% at 15~e$^-$. The selection uncertainties estimated through Monte Carlo are  negligible compared to that on fiducial volume, discussed later in Section~\ref{sec:background_fit}.

The final sample is shown in Fig.~\ref{fig:data_selection}. The S2-only sample is distributed in the energy region where most of the signal is expected, while the S1+S2 sample, distributed at the higher energies because of the S1 detection threshold, is key in constraining the background model discussed in the next section. The combined sample contains about 350,000 events and corresponds to a fiducial volume exposure,
including the spurious electron veto, of 12306~kg$\,$d.  The current  dataset  livetime is about 80\% larger than the one used in the 2018 analyses, which included data up to April 2017 \cite{DarkSide:2018bpj}

\section{Background model}
\label{sec:background}

The event rate in the energy range of interest for light dark matter search is dominated by $^{39}$Ar and $^{85}$Kr  decays,   originated in the LAr bulk, and by $\gamma$s and X-rays from radioactive isotopes in the detector components surrounding the active target.  The rate of NRs from radiogenic and cosmogenic neutrons and from  interactions of solar and atmospheric neutrinos, via coherent scattering off nucleus, is negligible with respect to the ER one,  and therefore not considered in this analysis.

\subsubsection{Internal  background}

$^{39}$Ar and $^{85}$Kr specific activities in underground argon were estimated in the first 70-days DarkSide-50 dataset     at  0.73$\pm$0.11~mBq/kg  and  2.05$\pm$0.13~mBq/kg, respectively, by fitting S1 spectra \cite{DarkSide:2015cqb}. The  $^{85}$Kr specific activity   was then corrected by its decay time ($\tau_{1/2}\sim$10.8~yr)  to  1.84$\pm$0.12~mBq/kg, %PA Jun22, Kr85 act scaled 
averaged over the  lifetime of the dataset used in this analysis. 

%PA Jun22, all the Kr85 act are scaled
$^{85}$Kr activity is  also assessed  by  identifying $\beta$-$\gamma$ fast coincidences from the 0.43\% decay branch to  $^{85m}$Rb  with  1.46~$\mu$s mean lifetime. This resulted in  an activity  of 1.82$\pm$0.15~mBq/kg, in excellent  agreement with the one obtained from the spectral fit. A third independent approach, based on the fit of the  $^{85}$Kr decay time  in the [50, 200]~$N_e$ range, resulted in a specific activity of 1.73$\pm$0.23~mBq/kg, compatible with the other two. The weighted mean of the three  measurements is 1.82$\pm$0.09~mBq/kg. 

\begin{figure}[h]
\includegraphics[width=1.0\linewidth]{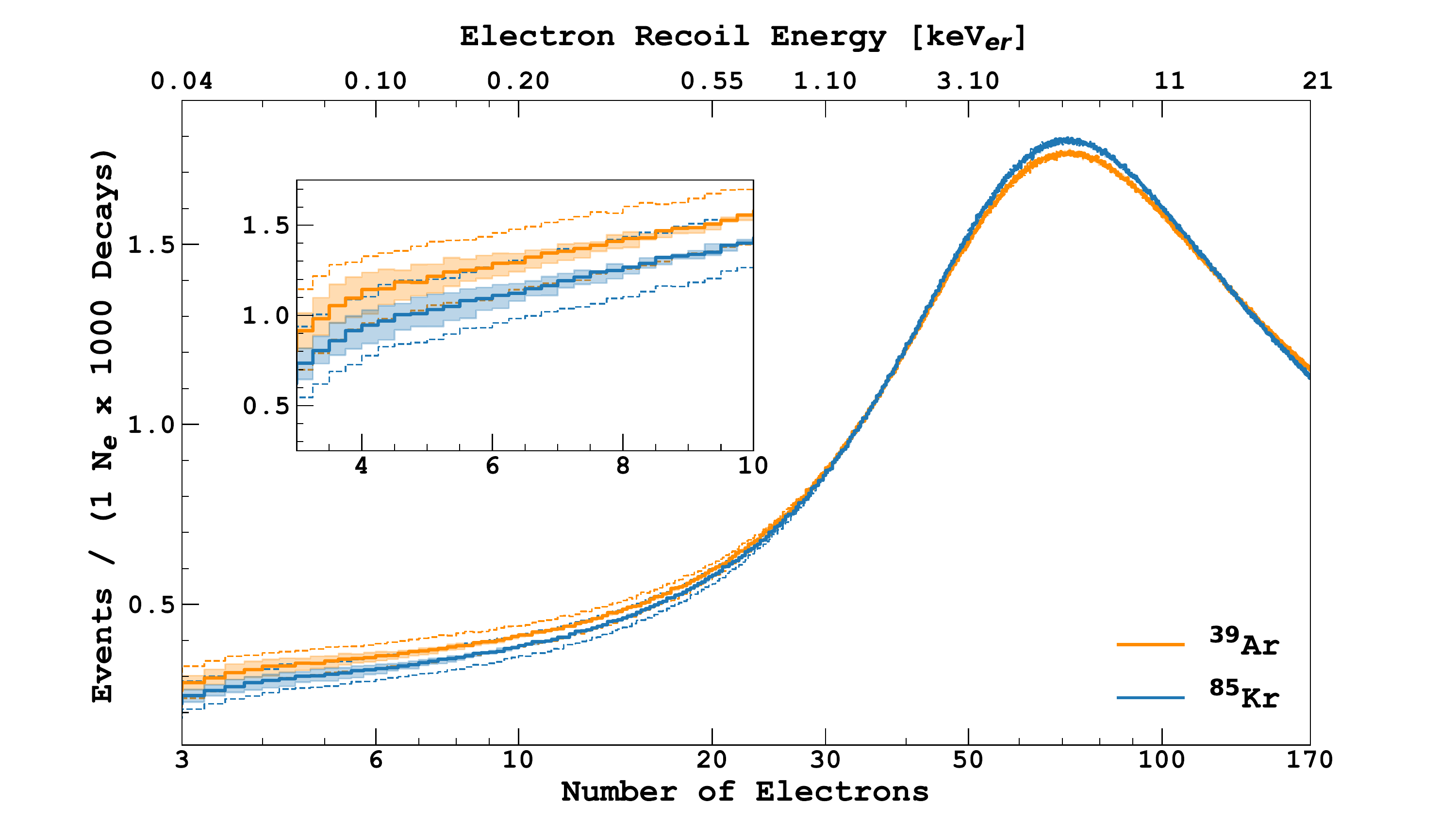}
\caption{$^{39}$Ar (orange) and $^{85}$Kr (blue) beta decay spectra in $N_e$ and associated systematics from atomic exchange and screening  effects (shaded area) and ionization response (dashed line). The systematic error propagated from the Q-value uncertainty  is too small to be illustrated  in this plot.}
\label{fig:arkr}
\end{figure}

%A major improvement in this analysis with respect to the one published in 2018 lies in the new spectral shapes
The  $^{39}$Ar and $^{85}$Kr ground-state to ground-state beta decays are first forbidden unique. The spectral shapes used in this analysis take into account recent calculations of atomic exchange and screening effects that have been extended to this transition nature \cite{PhysRevA.90.012501, Haselschwardt:2020iey},  validated  on measured $^{63}$Ni and $^{241}$Pu spectra with a 200~eV threshold.  Below this value, we assume a linear uncertainty on such corrections from 25\% at 0~eV up to 0\% at 200~eV. %\footnote{Recommended by the authors of ref.~\cite{PhysRevA.90.012501}  via private communication.}. 

Further systematics on the spectral shape originate from the uncertainty on the Q-value (1\% for $^{39}$Ar and 0.4\% for $^{85}$Kr), and from the detector response.  The main uncertainty on the latter,  as described in detail in ref.~\cite{DarkSide:2021bnz}, arises from the uncertainty on the ionization response, shown in Fig.~\ref{fig:qy}. 

The  $^{39}$Ar and $^{85}$Kr $N_e$-spectra, generated  including detector response effects,  are shown in Fig.~\ref{fig:arkr} together with the associated uncertainties.

The rates of $^{39}$Ar and $^{85}$Kr decay events falling inside the fiducial volume and in the energy of interest for this analysis are  evaluated in (6.5$\pm$0.9)$\times$10$^{-4}$~Hz and (1.7$\pm$0.1)$\times$10$^{-3}$~Hz, respectively, as reported in Table~\ref{tab:bg}. These are obtained using G4DS,  the Geant4-based DarkSide Monte Carlo toolkit \cite{DarkSide:2017wdu}.

\subsubsection{External  background}

\newcommand{\totalPMTActivityROI}{$(3.5 \pm 0.4)   \times 10^{-3} $} 
\newcommand{\totalCryoActivityROI}{$(5.9 \pm 0.4)  \times 10^{-4} $}

\begin{table}[h]\footnotesize
{\renewcommand{\arraystretch}{1.2}
\begin{tabular}{ ll|c| c| c }
  \hline
\multicolumn{2}{l|}{Location}    &  Activity    	& \multicolumn{2}{c}{Single-scatter events  in the RoI }     \\
\multicolumn{2}{l|}{and source} 	 & [Bq]                       &   Event rate  [Hz]  & Total rate [Hz]       \\
\hline
\multirow{2}{*}{\rotatebox[origin=c]{90}{LAr}} 
 & $^{39}$Ar & 0.034 $\pm{0.005}$   & $(6.5 \pm 0.9) \times 10^{-4} $  & $(6.5 \pm 0.9) \times 10^{-4} $\\  % \hline  % was .38. WHY?
 \cline{2-5} 
 & $^{85}$Kr  & 0.084  $\pm{0.004} $   & $(1.7 \pm 0.1) \times 10^{-3}$ & $(1.7 \pm 0.1) \times 10^{-3}$ \\  %updated Kr error (8.2%-->4.7% ) 
\hline
\multirow{12}{*}{\rotatebox[origin=c]{90}{PMT}}   
\multirow{6}{*}{\rotatebox[origin=c]{90}{Stems}}   
& $^{232}$Th   &  $0.16   \pm{0.03}$          &  $(3.2\pm 0.6)\times 10^{-4} $ &  \multirow{12}{*}{\totalPMTActivityROI } \\
& $^{238}$U up & $1.06   \pm{0.22}$         &  $(4.9\pm 1.0)\times 10^{-5} $ & \\
& $^{238}$U low& $0.34  \pm{0.03}$          &  $(3.2\pm 0.3)\times 10^{-4} $ & \\
& $^{235}$U  & $0.05  \pm{0.01}$            &  $(1.2\pm 0.2)\times 10^{-4} $ & \\
& $^{40}$K  & $2.39 \pm{0.32}$              &  $(1.8\pm 0.2)\times 10^{-4} $ & \\
& $^{54}$Mn & $0.05   \pm{0.02}$            &  $(3.5\pm 1.4)\times 10^{-5}$ & \\  
\cline{2-4} 
\multirow{5}{*}{\rotatebox[origin=c]{90}{}}   
\multirow{5}{*}{\rotatebox[origin=c]{90}{}}   
\multirow{5}{*}{\rotatebox[origin=c]{90}{}}   
\multirow{5}{*}{\rotatebox[origin=c]{90}{Ceramic}}   
& $^{232}$Th    & $0.07   \pm{0.01}$     &  $(2.4\pm 0.3)\times 10^{-4} $& \\
& $^{238}$U up  & $4.22   \pm{0.88}$     &  $(4.2\pm 0.9)\times 10^{-4} $& \\
& $^{238}$U low & $0.34   \pm{0.03}$     &  $(5.3\pm 0.5)\times 10^{-4} $& \\
& $^{235}$U     & $0.21   \pm{0.03}$     &  $(9.8\pm 1.4)\times 10^{-4} $& \\
& $^{40}$K      & $0.61  \pm{0.08}$      &  $(8.1\pm 1.1)\times 10^{-5} $& \\

\cline{2-4} 
\multirow{5}{*}{\rotatebox[origin=c]{90}{}}   
\multirow{5}{*}{\rotatebox[origin=c]{90}{}}   
\multirow{5}{*}{\rotatebox[origin=c]{90}{}}   
\multirow{2}{*}{\rotatebox[origin=c]{90}{Body}} & \multirow{2}{*}{$^{60}$Co}  & \multirow{2}{*}{$0.17   \pm{0.02}$}     &  \multirow{2}{*}{$(2.4 \pm 0.3)  \times 10^{-4} $} & \\   
&   &      &    & \\
\hline 
\multirow{6}{*}{\rotatebox[origin=c]{90}{Cryostat}}    
& $^{232}$Th  & $ 0.19  \pm{0.04}$            & $(7.9\pm 1.7)\times 10^{-5} $& \multirow{6}{*}{\totalCryoActivityROI } \\
&$^{238}$U up   & $ 1.30  \pm{0.2}$           & $(1.5\pm 0.2)\times 10^{-5} $&\\
&$^{238}$U low  & $ 0.38  ^{+0.04}_{-0.19}$   & $(5.3 ^{+0.6}_{-2.6})\times 10^{-6} $&\\
&$^{235}$U      & $ 0.045 ^{+0.01}_{-0.02}$   & $(1.5 ^{+0.3}_{-0.7})\times 10^{-5} $&\\ 
&$^{60}$Co      & $ 1.38  \pm{0.1}$           & $(4.7\pm 0.3)\times 10^{-4} $&\\
&$^{40}$K &     $ 0.16  ^{+0.02}_{-0.05}$     & $(3.4 ^{+0.4}_{-1.1})\times 10^{-6} $&\\
\hline
\end{tabular}
\caption{Background activities and event rate in the RoI from the bulk, PMTs, and cryostat from material screening.  The activity measurements are reported for chain progenitors only, while the event rates are quoted for full decay chains. The uncertainties are propagated from the screening measurements. An additional 10\% systematic error is included in the PMT error, due to the uncertainty on the contamination partitioning between stems and body. }
\label{tab:bg}
}
\end{table}

This analysis is based on extensive simulation of each external background component as measured in the  material screening campaign, unlike the 2018 analyses where the external background model was extrapolated from high-energy fits.

The main sources of external X-ray and $\gamma$ background are the  radioactive contaminants in  PMTs and in the stainless-steel cryostat,  both characterized with an extensive materials assay campaign. These measurements and their associated errors, once corrected for decreased activity due to elapsed time at the dataset date, are the inputs for the external background model.

The PMT components,  which dominate in terms of radioactivity, are the  stems in the back of the PMT, the ceramic around the dynodes chain, and the  PMT body made of kovar.  In contrast, the cryostat is composed only of stainless steel, where contamination is uniformly distributed.

The radioactive isotopes measured in the screening campaign are quoted in Table~\ref{tab:bg}. The secular equilibrium of $^{238}$U is broken at the level of $^{222}$Rn, because different activities were observed between the top and bottom of the chain. Each isotope is simulated uniformly distributed in the component material, and decaying particles are tracked over all the DarkSide-50 geometry with G4DS.   The detector response to each deposit is simulated with the Monte Carlo strategy, which includes, among others, electron captures by impurities along the drift and the dependence of the S2 response on the event radial position~\cite{DarkSide:2021bnz}.  Energy dependent inefficiencies from the quality cuts are applied to the Monte Carlo sample on an event-by-event basis.

\begin{figure}
\centering
\includegraphics[width=1.0\linewidth]{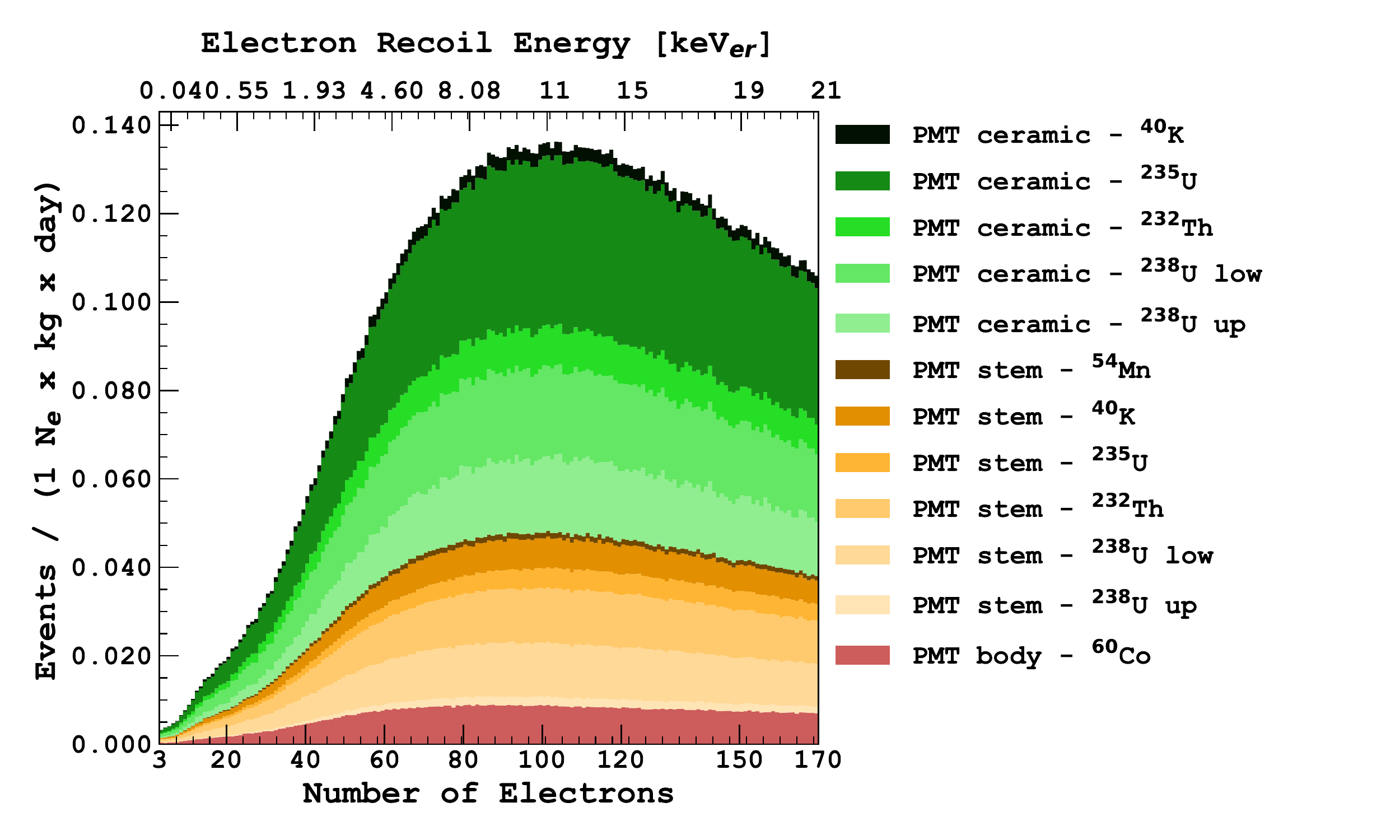}
\includegraphics[width=1.0\linewidth]{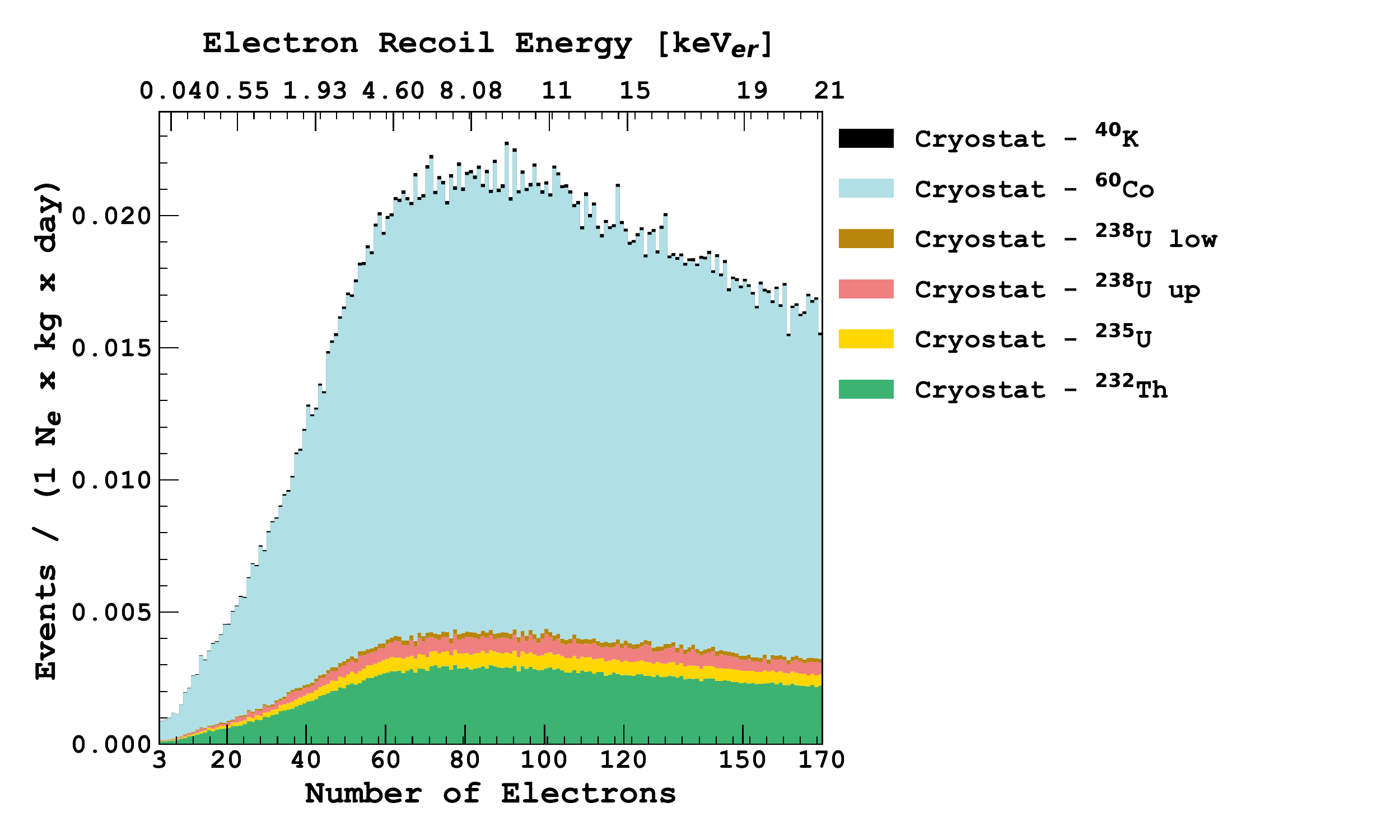}
\caption{Breakdown of  PMTs (top)  and cryostat (bottom) radioactive contributions. The stacked spectra in the energy region of interest and in the fiducial volume are scaled by the measured activity.}
\label{fig:ext_bgs}
\end{figure}

Table  \ref{tab:bg} reports  activities and  errors from the material screening, and the event rate of single scatter events, within the fiducial volume and in the energy region of interest, for each contribution.  The resulting predicted event rates for PMTs and cryostat components   are \totalPMTActivityROI~Hz    and   \totalCryoActivityROI~Hz, respectively. The difference in the source location has a large impact on the induced event rate, due  to the distance of the component    from the active volume.  Therefore, an additional 10\% systematic error, derived from Monte Carlo simulations,  is accounted for the PMTs contribution due to the uncertainty on the contamination partitioning    between stems and body. The resulting energy spectra  with the breakdown of the radioactive contributions for both PMTs and cryostat are shown in Fig.~\ref{fig:ext_bgs}. The cryostat component is largely dominated by $^{60}$Co, and the PMTs  one by the contamination from the ceramic. 

It is worth noting that the spectra from individual contributions for each of the two components are nearly indistinguishable, thus reducing the impact of systematics related to their individual normalizations. We verified that the systematics related to possible spectral deformation is negligible by comparing the summed spectra while varying the amplitude of each component by 1-$\sigma$, as quoted in Table~\ref{tab:bg}.

\begin{figure}
\centering
\includegraphics[width=1.0\linewidth]{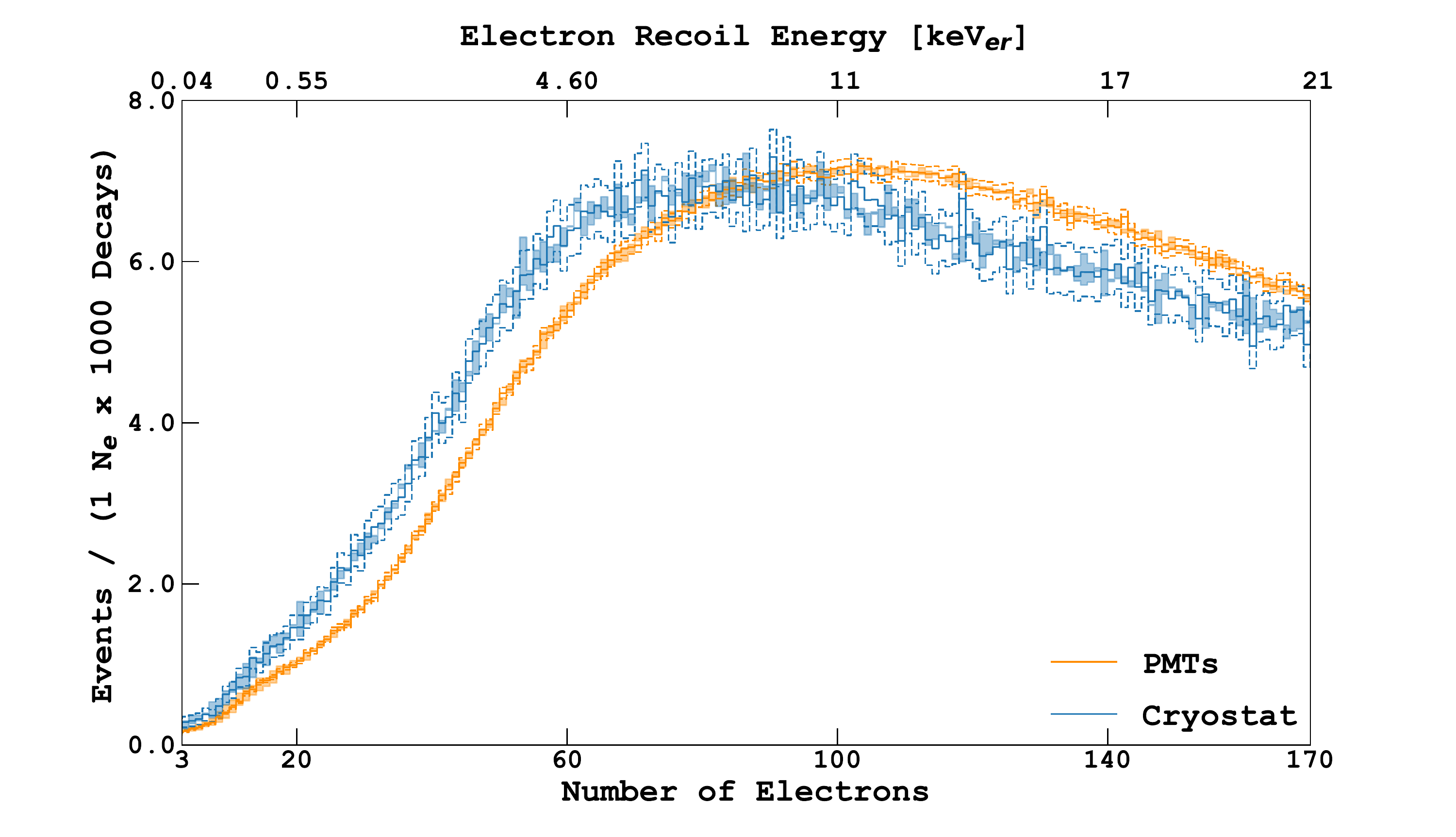}
\caption{Background spectra from PMTs and cryostat and associated error bands from the detector response (shaded area) and from the Monte Carlo statistics (dotted lines).}
\label{fig:ext_sys}
\end{figure}

The final spectra are shown in Fig.~\ref{fig:ext_sys}, together with the systematic error from the detector response and the Monte Carlo statistical  uncertainty.

\section{Background-only fit}
\label{sec:background_fit}

The analysis is based on a binned  profile likelihood, $\mathcal{L}$, implemented through the RooFit/HistFactory package~\cite{Cranmer:1456844},

 \begin{eqnarray}
  \mathcal{L}&  =  & \prod_{i\,\epsilon\,\textrm{bins}} \mathcal{P} \left( n_i| m_i(\mu_s,\Theta)\right) \times  \prod_{\theta_i\,\epsilon\, \Theta}\mathcal{G}(\theta_i|\theta^0_i\textrm{,}\,\Delta \theta_i) \nonumber \\
   &\times &  \prod_{i\,\epsilon\,\textrm{bins}}  \mathcal{G}\left( m_i(\mu_s,\Theta)| m^0_i(\mu_s,\Theta), \delta m_i(\mu_s,\Theta)\right). 
 \label{eq:likelihood}
 \end{eqnarray}

The first term represents the Poisson probability of observing $n_i$ events in the $i^{th}$-bin with respect to the expected ones, $m_i(\mu_s, \Theta)$, with $\mu_s$ the signal   strength  and  $\Theta$ the set of nuisance parameters. 
 The second term includes the  Gaussian penalty terms to account for  the nuisance parameters ($\theta_i^0$ and $\Delta\theta_i$ are the nominal central values and uncertainties, respectively), listed in  Table~\ref{tab:sys}, which may act   on multiple components in a correlated way. The last term of eq.~\ref{eq:likelihood} accounts for the statistical uncertainties in each bin    ($m_i(\mu_s,\Theta)\pm\delta m_i(\mu_s,\Theta)$) of the simulated sample with respect to the nominal value, $m^0_i(\mu_s,\Theta)$~\cite{Cranmer:1456844}. 
 
Nuisance parameters are classified as ``amplitude'' parameters, acting on the   normalization  of the background components, or as ``shape'' parameters, accounting for  spectral distortions from the ionization response and from  uncertainties on  $^{39}$Ar and $^{85}$Kr $\beta$-decays. Among the amplitude parameters, the uncertainty on fiducial volume due to thermal contraction of PTFE has a different impact on internal and external components. From one side, the relative uncertainty on the activities of $^{39}$Ar and $^{85}$Kr decays, uniformly distributed in the TPC,  is equal to the fiducial volume one (1.5\%). On the other hand, the fraction of the external background falling inside the fiducial volume depends on the positions of the PMTs, which are installed on a PTFE holder and thus subjected to thermal contraction. This uncertainty has been propagated with Monte Carlo simulations and results equal to 1.1\% for both the cryostat and PMTs backgrounds. 

As for the shape systematics, these are implemented through a template morphing based on a vertical bin interpolation between  histograms  distorted by systematics. This approach also allows for accounting for asymmetric errors \cite{Barlow:1993dm}.

\begin{table*}
%\begin{xtabular*){lll}
%\begin{tabular}{l|p{4cm}|p{2.5cm} }
\begin{tabular}{l|l|l|l }
%\begin{tabularx}{1.0\textwidth}{ l l l l }
\hline
  & Name &   Source  & Affected components      \\
\hline
\multirow{5}{*}{\rotatebox[origin=c]{90}{Amplitude}}    
& A$_{FV}$  &  uncertainty on the fiducial volume & WIMP, $^{39}$Ar, $^{85}$Kr,  PMTs, Cryostat  \\ \cline{2-4} 
\cline{2-4} 
& A$_{Ar}$   & 14.0\% uncertainty on $^{39}$Ar activity  & $^{39}$Ar \\
\cline{2-4} 
& A$_{Kr}$   & 4.7\% uncertainty on $^{85}$Kr activity & $^{85}$Kr \\ %updated Kr error (8.2%-->4.7% )
\cline{2-4} 
& A$_{pmt}$    & 11.5\% uncertainty on  activity from PMTs  & PMT \\
\cline{2-4} 
& A$_{cryo}$  & 6.6\%  uncertainty on  activity from the cryostat    & Cryostat \\
\hline 
\multirow{6}{*}{\rotatebox[origin=c]{90}{Shape}}    
& Q$_{Kr}$  &   0.4\% uncertainty on the $^{85}$Kr-decay Q-value & $^{85}$Kr\\
\cline{2-4} 
& Q$_{Ar}$   &   1\%   uncertainty on the $^{39}$Ar-decay Q-value  &  $^{39}$Ar \\
\cline{2-4} 
& S$_{Kr}$  &  spectral shape uncertainty on atomic exchange and  screening effects & $^{85}$Kr \\
\cline{2-4} 
& S$_{Ar}$ &  spectral shape uncertainty on atomic exchange and  screening effects & $^{39}$Ar \\
\cline{2-4} 
&  $Q_y^{er}$ & spectral shape  systematics from  ER ionization response uncertainty   &  $^{39}$Ar, $^{85}$Kr, PMTs, Cryostat \\
\cline{2-4} 
 & $Q_y^{nr}$ & spectral shape  systematics from  NR ionization response uncertainty   & WIMP   \\
\hline 
\end{tabular}
\caption{List of systematics, their sources, and impacted signal and background components included in the binned profile likelihood. Any considered signal is equally affected by the uncertainty on the dataset exposure, but differs on the ionization response, on the basis of the recoil type. WIMP-nucleon interactions are subjected to the NR ionization response uncertainty.  }
\label{tab:sys}
\end{table*}

\begin{figure}[h]
\begin{center}
\includegraphics[width=1.0\linewidth]{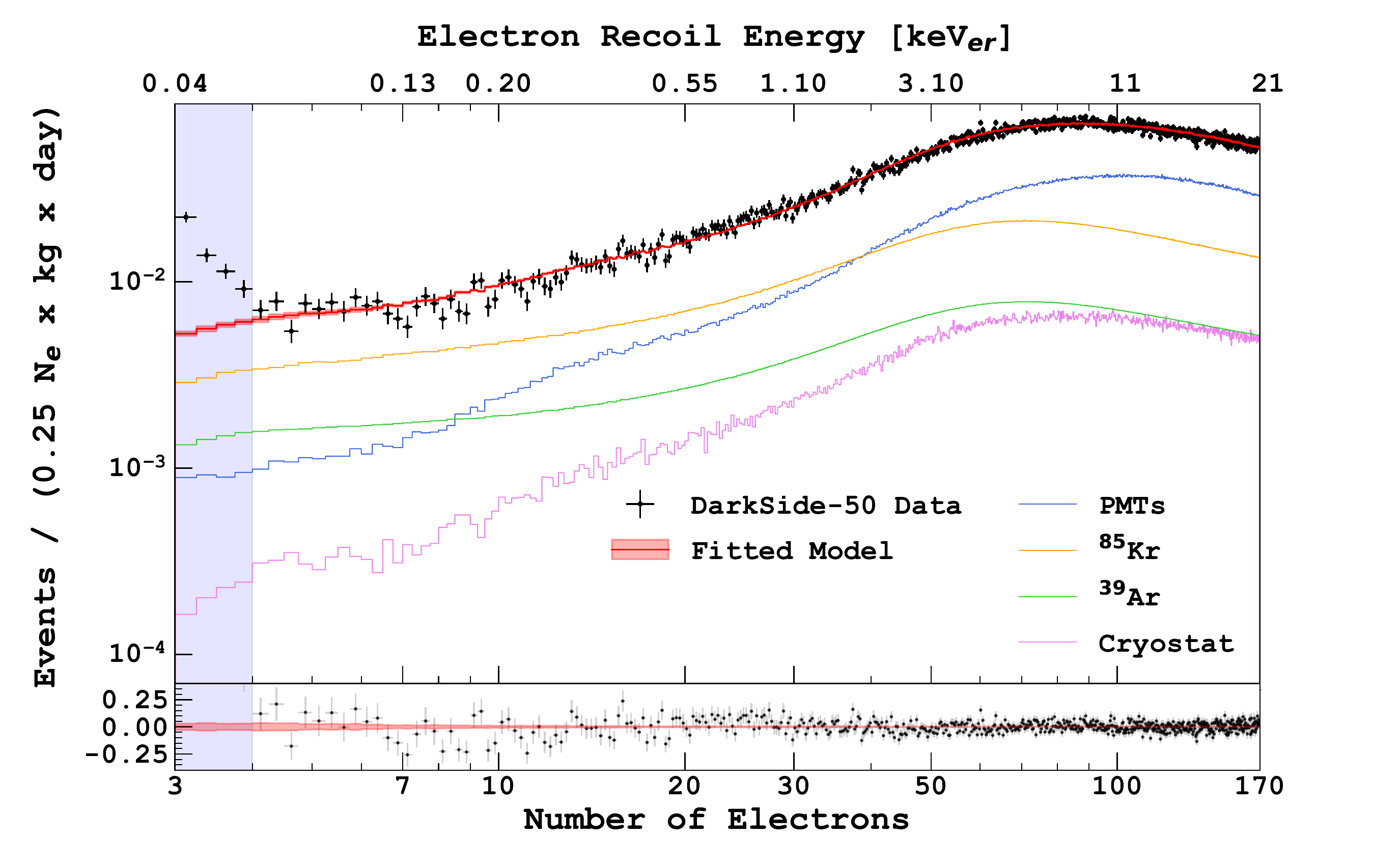}
\caption{Background model and uncertainty (red line and shaded area) from the data fit  in the [4, 170]~$N_e$ range, and the individual contributions from the internal ($^{39}$Ar and $^{85}$Kr) and external components (cryostat and PMTs).  %The result does not change when narrowing the fit range to [7, 170]~$N_e$ (blue line). 
An excess of events  with respect to the background model is observed below 4~$N_e$ (blue shaded area). The residuals, defined as the difference between the observed and expected events, normalized to the expected ones, are compared below to the model uncertainty from the fit. }
\label{fig:fit_from_4e}
\end{center}
\end{figure}

Fitting  the data with  background only components, by removing the signal component   from eq.~\ref{eq:likelihood}  (see Fig.~\ref{fig:fit_from_4e}), allows for model diagnostics.  An excess of events was observed between 4 and 7~$N_e$ in the 2018 analyses~\cite{DarkSide:2018bpj,DarkSide:2018ppu}. In the current analysis, the fit from  4~$N_e$ is compatible with data, confirming the suppression of the event excess.  There are multiple reasons for this improvement: the more efficient background rejection of quality cuts, the better modelling of $^{39}$Ar and $^{85}$Kr $\beta$-spectra, including   atomic exchange  and screening effects,  particularly important at low energies, and a more accurate treatment of systematics. 

\begin{figure}[h]
\begin{center}
\includegraphics[width=1.0\linewidth]{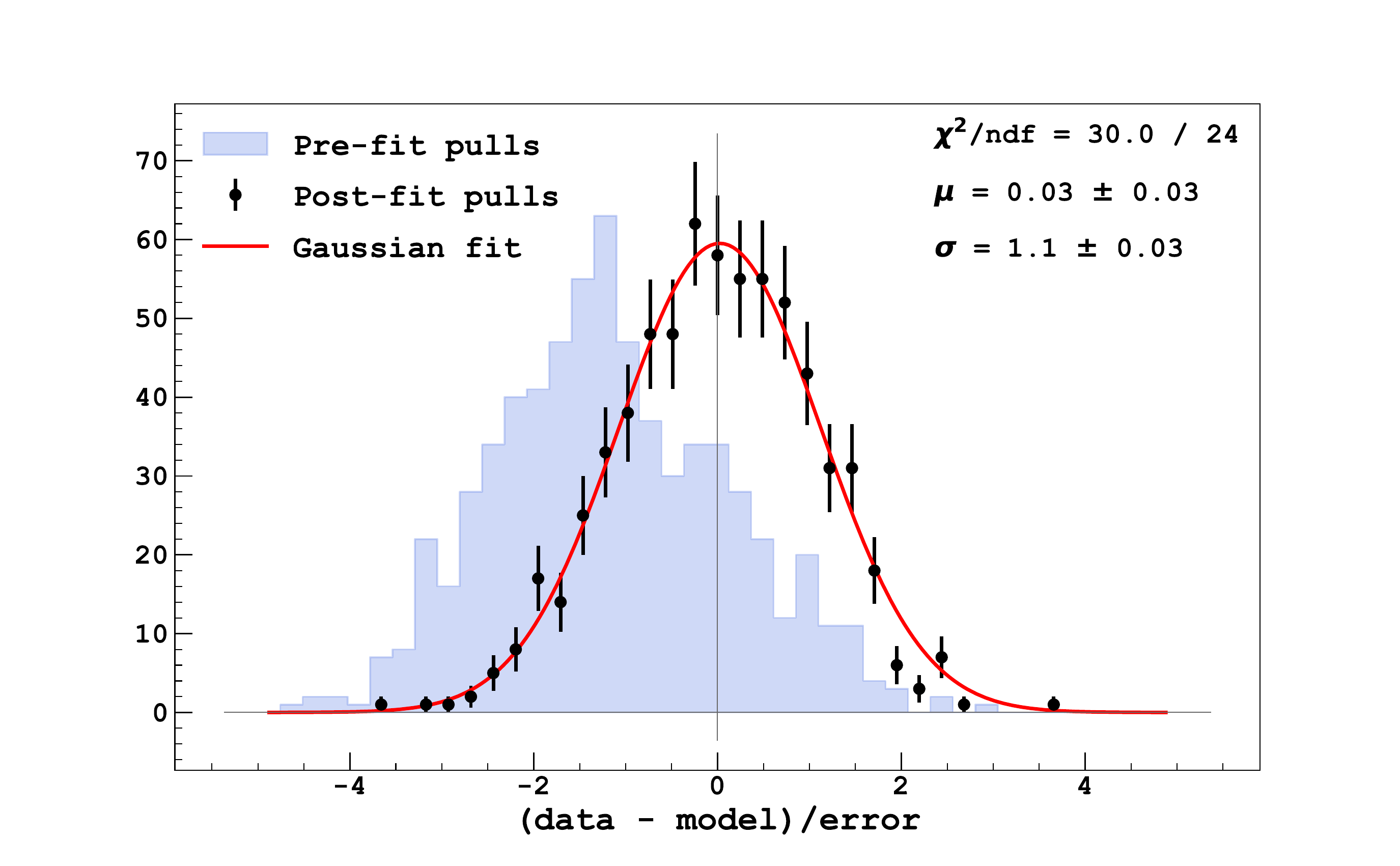}
\caption{Pulls from the background-only fit  (black points) are normally distributed, as highlighted by the Gaussian fit  (red line). Shaded blue histogram corresponds to pre-fit distribution.}
\label{fig:pulls}
\end{center}
\end{figure}

\begin{figure}[h]
\begin{center}
\includegraphics[width=1.0\linewidth]{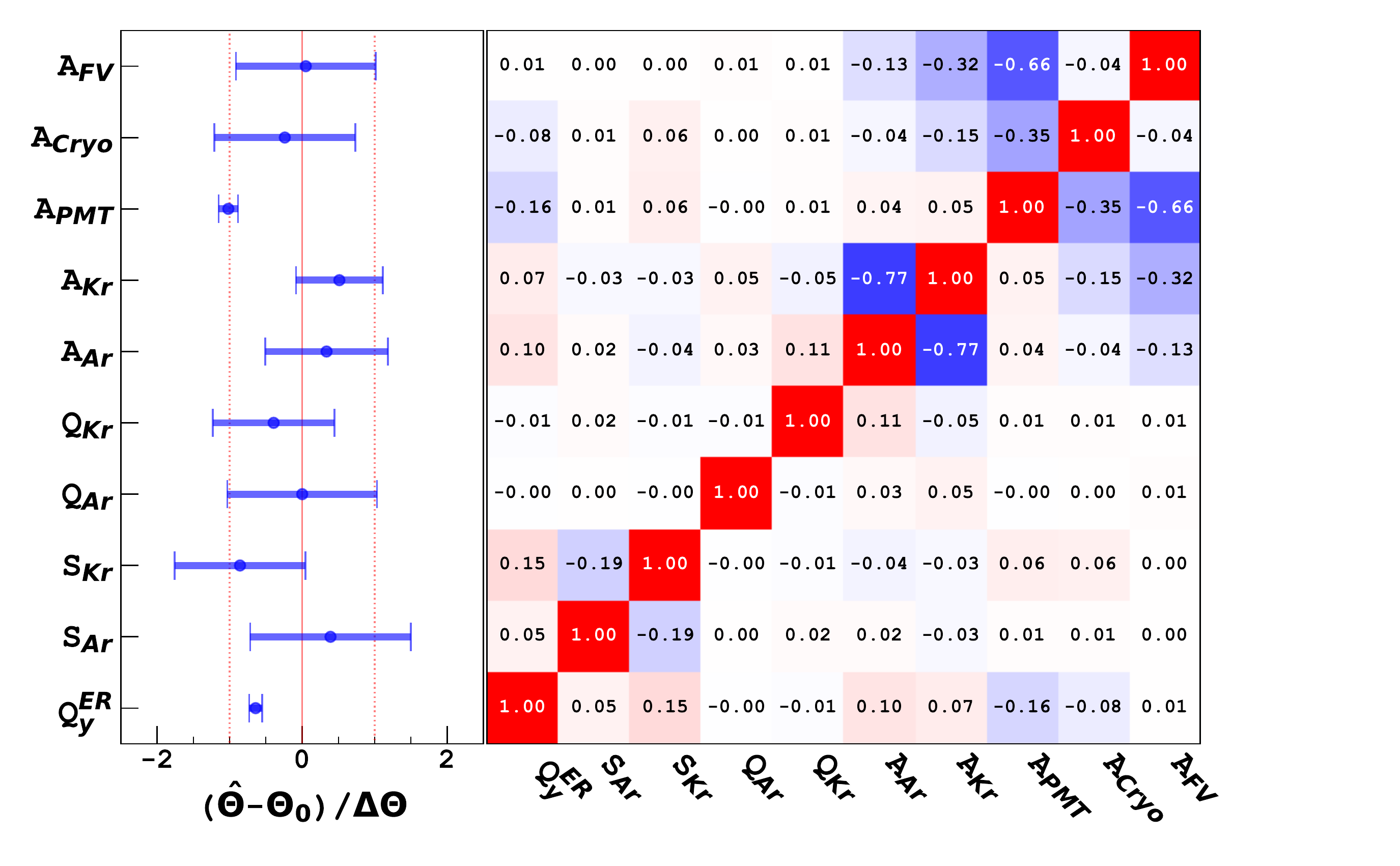}
\caption{Post-fit nuisance parameters  compared to the nominal values (left) and  correlation matrix (right) from the background-only fit. Error bars are normalized to the pre-fit size of each of the nuisance parameter penalty terms. }
\label{fig:correlation}
\end{center}
\end{figure}

The  data pulls from the  fit  in the full [4, 170]~$N_e$ range are normally distributed, as shown in Fig.~\ref{fig:pulls}, demonstrating  the quality of the background model. Post-fit values of nuisance parameters  are in good agreement with  the nominal ones,  as reported  in Fig~\ref{fig:correlation}. Improvement between nominal and post-fit   errors  are observed for two parameters: the amplitude of the PMT component and the ionization response to ERs. The former is subject to a large uncertainty from the position of the contaminant within the PMT itself, as discussed in  Sec.~\ref{sec:background}, and the latter relies on a few calibration points, especially at low energies~\cite{DarkSide:2021bnz}. In both cases, the fitted dataset provides additional information to improve their uncertainty, as also observed from the fit  of an Asimov sample derived from the background model~\cite{Cowan:2010js}. 

As shown in Fig~\ref{fig:correlation}, the similarity in spectra between various components, as for $^{39}$Ar and $^{85}$Kr or PMTs and cryostat backgrounds,  is at the origin of their anti-correlation. The impact of fiducial volume uncertainty, which  equally acts  on all normalization factors, is larger when relative uncertainties on component amplitudes are higher.

Finally,  we tested the hypothesis of tritium contamination from cosmogenic activation during the LAr transportation, which was estimated of the order of a few 0.1~mBq/kg. However, hydrogen is chemically removed by the hot getter present in the DarkSide-50 gaseous purification loop, hence we expect a negligible tritium contamination in our target. Notice that tritium is also one of the hypotheses to explain the excess of events observed by Xenon1T \cite{XENON:2020rca}. We probed its presence in DarkSide-50 by adding an unconstrained component in the likelihood. The tritium activity in DarkSide-50 was found to be compatible with zero, with an upper limit of  $<$1.1$\times$10$^{-3}$~mBq/kg at 90\% C.L. ($<$2$\times 10^{-5}$~Hz in the fiducial volume and in the RoI), and thus not included in the analysis.

\section{Sensitivity to WIMP-nucleon interactions}

The signal from spin-independent WIMP-nucleon scattering is derived assuming the standard isothermal WIMP halo model, with $v_{esc}\,$=$\,$544~km/s, $v_0\,$=$\,$238~km/s, $v_{Earth}\,$=$\,$232~km/s, and $\rho_{DM}\,$=$\,$0.3~GeV/c$^2$/cm$^3$ \cite{Smith:2006ym, Baxter:2021pqo}.

\begin{figure}[h]
\begin{center}
\includegraphics[width=1.0\linewidth]{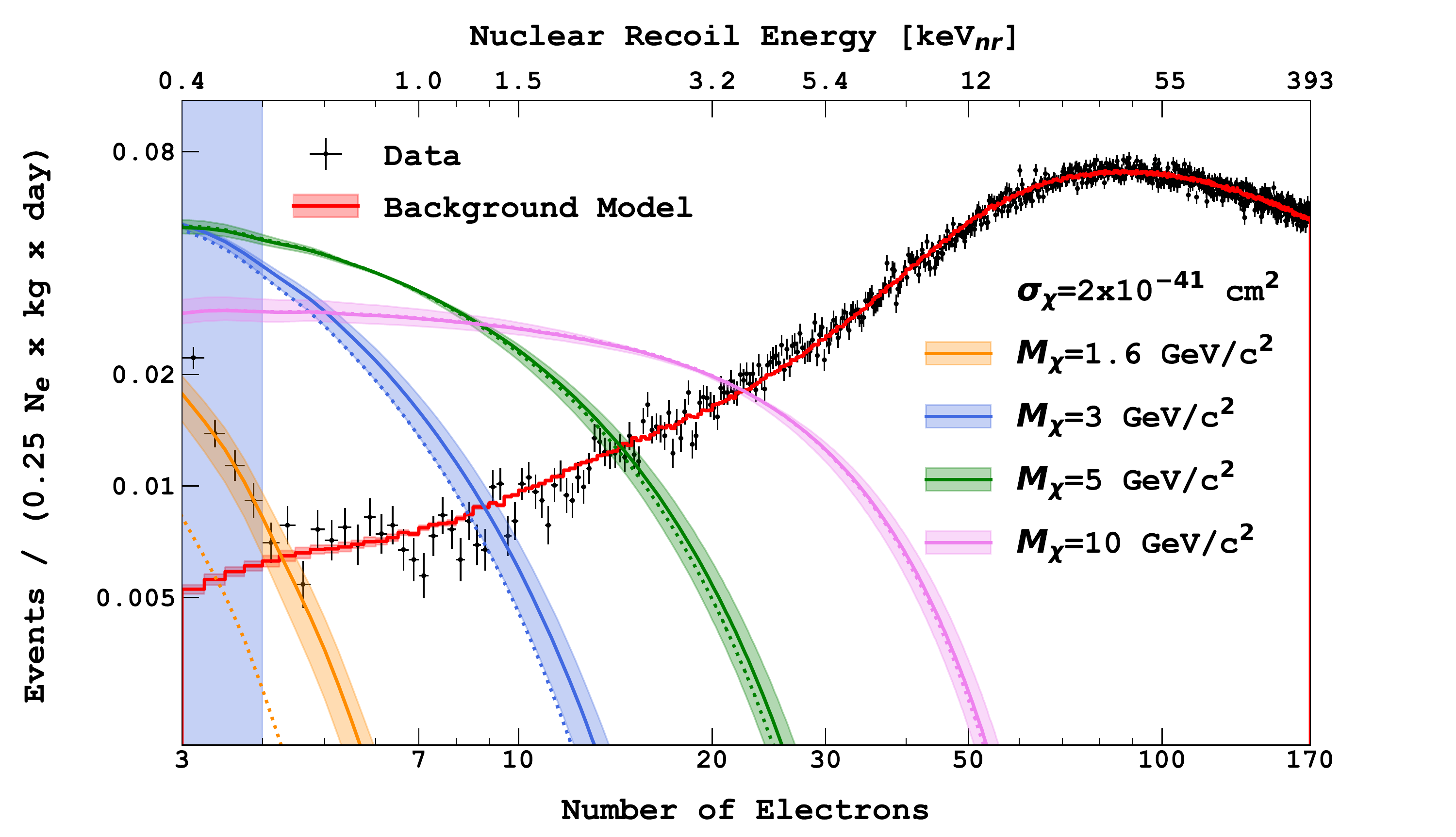}
\caption{Data and background model compared to expected WIMP spectra, assuming binomial quenching fluctuations (solid lines) and  WIMP-nucleon scattering cross section  equal to  2$\times$10$^{-41}$~cm$^2$.  The systematic error associated with WIMP spectra is due to uncertainty on the NR ionization response. For reference, WIMP spectra assuming no quenching fluctuation (dashed lines) are also shown. }
\label{fig:signal}
\end{center}
\end{figure}

Recoils from WIMPs, via elastic scattering off nucleons, are modeled with a Monte Carlo approach, as done for the background components (see Sec.~\ref{sec:background}) and including the ionization response to NRs shown in Fig~\ref{fig:qy}. The main unknown in such response is the fluctuation from the ionization quenching effect, an issue already raised in the 2018 analysis \cite{DarkSide:2018bpj} and still unresolved. Quenching fluctuations, in addition to fluctuations resulting from the partitioning between excitons and ionization electrons and from ion-electron recombinations, play a key role because, at very low energies, they increase the probability of observing events above the analysis threshold. The suppression of quenching fluctuations, though not physical, represents  the most conservative modelling with respect to the WIMP search.  An alternative model, also considered in this analysis, relies on binomial quenching fluctuations, \textit{i.e.}, between detectable (ionization electrons and excitons) and undetectable quanta (e.g. phonons). The choice of  binomial fluctuations ensures that the number of produced quanta  does not exceed the maximum one,  equivalent to the ratio between the deposited energy and the average work function (19.5~eV~\cite{Doke:2002oab}) in LAr \cite{DarkSide:2021bnz}. The comparison between expected WIMP signals, assuming quenching fluctuations,  and background model and data, is shown in Fig~\ref{fig:signal}.

\begin{figure}[h]
\begin{center}
\includegraphics[width=1.0\linewidth]{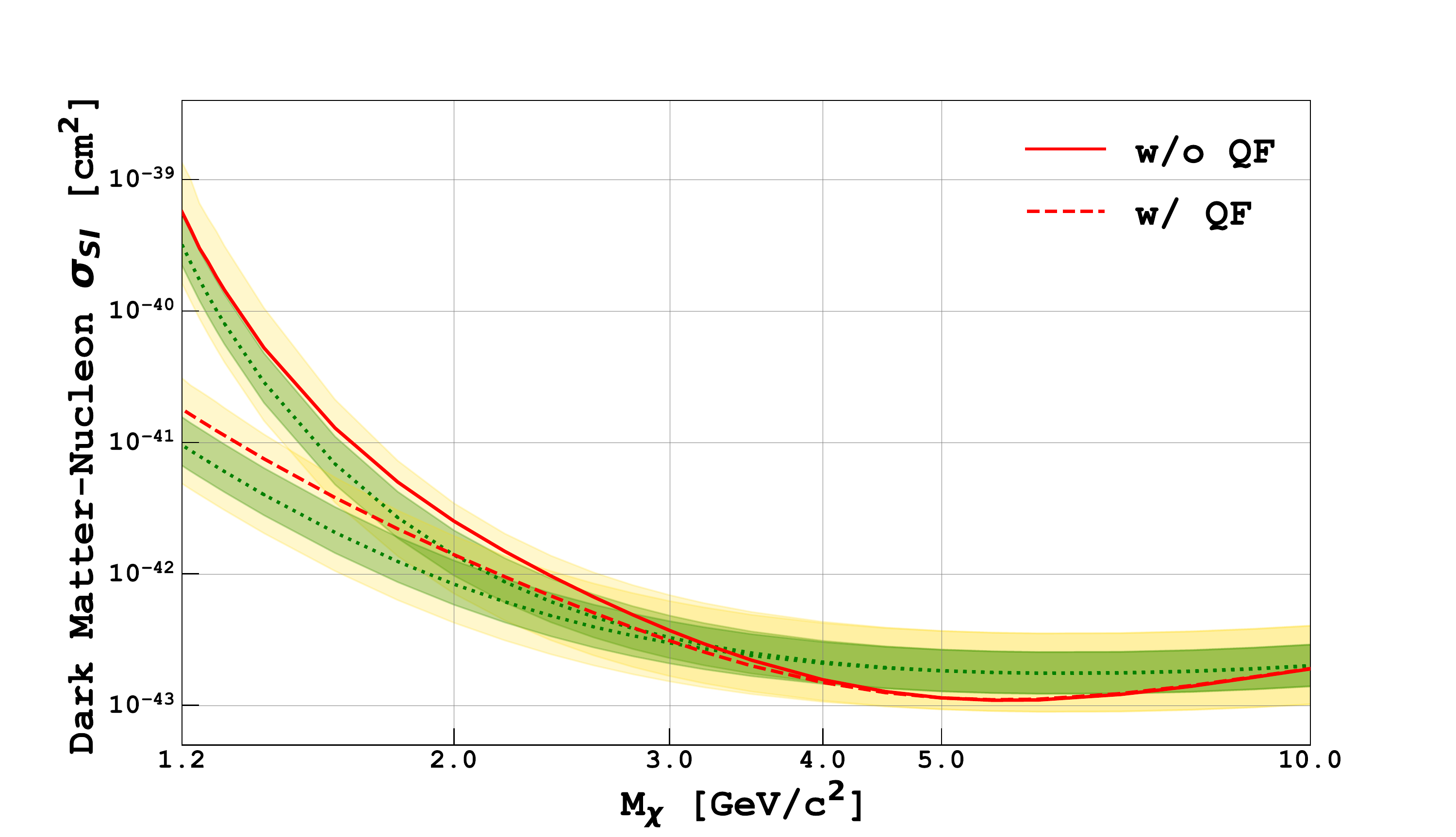}
\caption{90\% upper limits on spin independent WIMP- nucleon cross sections from DarkSide-50 in the range above 1.2~GeV/c$^{2}$. Both non-quenching (NQ, solid red line) and  quenching (QF, dashed red line) fluctuations models  are considered. Also shown are the expected limits (green dotted lines) with the $\pm$1-$\sigma$ (green shaded area) and $\pm$2-$\sigma$ (yellow shaded area) bands.}
\label{fig:wimp_limit_exp}
\end{center}
\end{figure}

The nuisance parameters considered in eq.~\ref{eq:likelihood}, include those already discussed for the background-only fit (see Sec.~\ref{sec:background_fit} and Table~\ref{tab:sys}), as well as those associated to the signal. The dominant one is  the uncertainty on the NR ionization response, shown in Fig.~\ref{fig:qy}, obtained from the simultaneous fit \cite{DarkSide:2021bnz} of spectra from calibration neutron sources deployed in the DarkSide-50 veto, and external calibration datasets from  SCENE~\cite{SCENE:2014iyj} and ARIS~\cite{Agnes:2018mvl} test beam experiments. The signal amplitude is also affected by the uncertainty on the fiducial volume, a systematic correlated with the background components. 

The observed upper limit of 90\% C.L. computed with the CLs technique \cite{Read:2002hq} for the two signal models, with (QF) and without (NQ) quenching fluctuations, are shown in Fig.~\ref{fig:wimp_limit_exp}, along with the expected limits. In both cases,  observed limits are compatible within 2-$\sigma$ with the expected ones, and coincide between them above 4~GeV/c$^2$ WIMP mass, where the impact of quenching fluctuations is negligible. 

\begin{figure}[h]
\begin{center}
\includegraphics[width=1.0\linewidth]{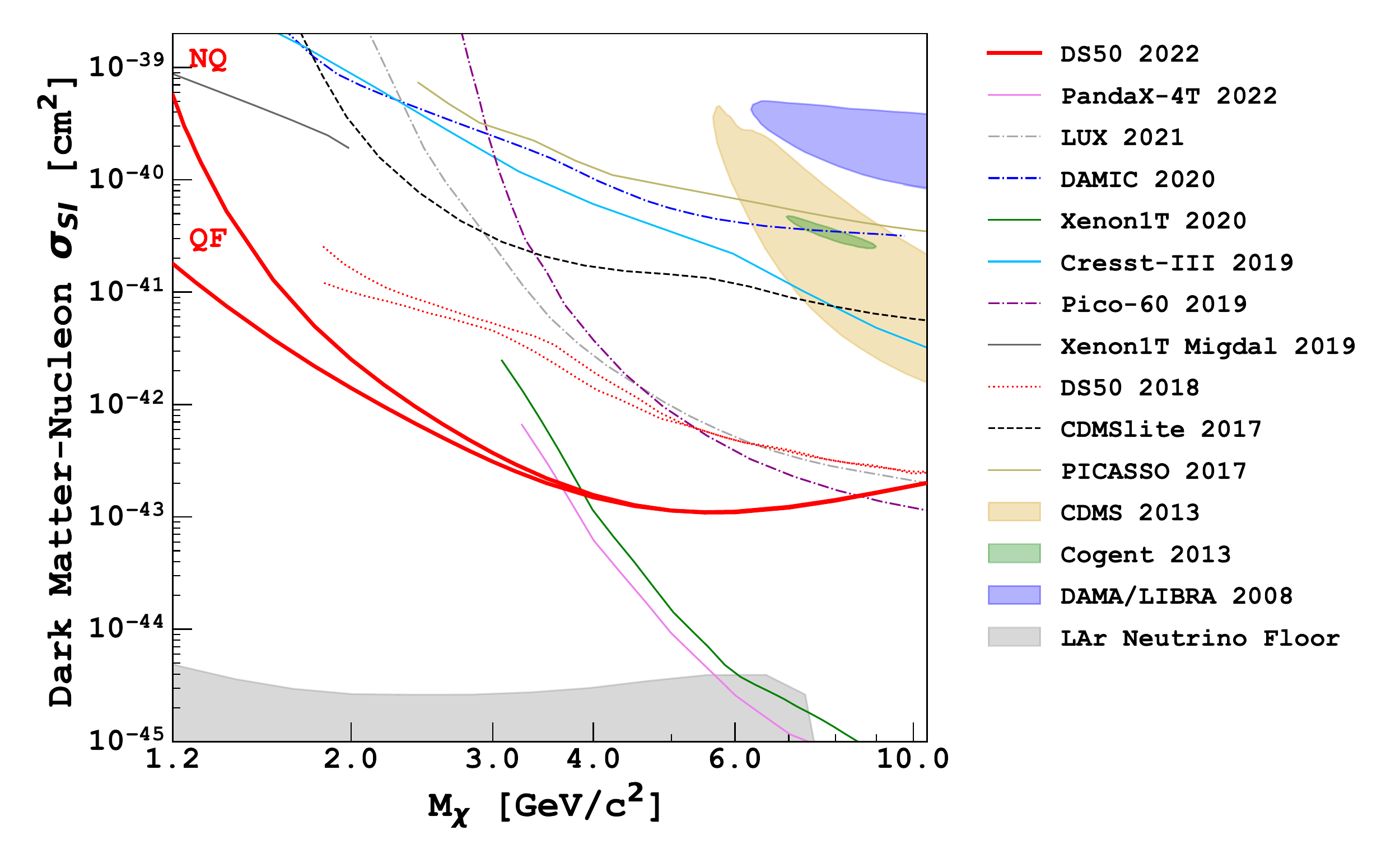}
\caption{DarkSide-50 limits with and without quenching
fluctuations are labeled QF and NQ, respectively. These limits are compared to the 90\% C.L. exclusion limits and claimed discovery from Refs.~\cite{Ma:2022wdz,DAMIC:2021esz, CRESST:2019jnq, Angloher:2011uu,PICO:2017tgi,XENON:2019zpr, XENON:2020gfr, DarkSide:2018bpj, SuperCDMS:2017nns, SuperCDMS:2013eoh, LUX:2020yym, Behnke:2016lsk, CoGeNT:2010ols, Bernabei:2013xsa} and to the neutrino floor for LAr experiments  \cite{Ruppin:2014bra}.}
\label{fig:wimp_limit}
\end{center}
\end{figure}

Exclusion limits above 1.2~GeV/c$^2$ are compared  in Fig.~\ref{fig:wimp_limit} with the 90\% C.L. exclusion limits and with regions of claimed discovery  from Refs~\cite{PandaX-4T:2021bab,DAMIC:2021esz, CRESST:2019jnq, PICO:2017tgi,XENON:2019zpr,XENON:2019gfn, XENON:2018voc, DarkSide:2018bpj, SuperCDMS:2017nns, SuperCDMS:2013eoh, LUX:2016ggv, Behnke:2016lsk, CoGeNT:2010ols, Bernabei:2013xsa}. Assuming non-quenching fluctuations, the most conservative model, DarkSide-50 establishes, with this work, the world's best limits for WIMPs with masses in the range [1.2, 3.6]~GeV/c$^2$ and improves on the previous one by a factor of $\sim$10 at 3~GeV/c$^2$. The dominant factors that enabled this improvement are the data selection, which suppressed the excess observed in 2018 over the background model in the region between 4 and 7~$N_e$~\cite{DarkSide:2018bpj}, and the inclusion of atomic corrections to the spectra of first unique forbidden $^{39}$Ar and $^{85}$Kr decays,  which improve the agreement  between data and model.

The DarkSide-50 limits described in this work are confirmed using an alternative Bayesian approach, where the analytical ER and NR calibration responses are made explicit  in the likelihood. This approach allows  to propagate systematic uncertainties in the final result without any intermediate Gaussian or linearity assumptions.  In addition, the likelihood is marginalized, and not profiled with respect to the nuisance parameters, as done in this work. This yields a reliable estimate of uncertainties even when a multivariate normal distribution does not provide a good approximation of the likelihood function. The description of this approach and of associated DarkSide-50 results will be released soon in a dedicated publication.

\section{Conclusions}
In this work, we reanalyzed the DarkSide-50 dataset used in 2018 \cite{DarkSide:2018bpj} to produce the exclusion limit in the region below 10~GeV/c$^2$. Compared to the previous analysis, we improved the data selection, calibration of the detector response, background model, and determination of systematic errors. The good understanding of the background down to 0.6 keV$_{nr}$, corresponding to 4 electrons, allows  to improve the previous DarkSide-50 exclusion limit by a factor of about 10 at 3~GeV/c$^2$. More generally, this analysis has produced the world's best limit on the spin-independent WIMP-nucleon elastic scattering in the region between 1.2 and 3.6~GeV/c$^2$, assuming the signal model without quenching fluctuations, i.e. the most conservative hypothesis. The same analysis approach was also applied  to  improve existing limits on spin-independent WIMP-nucleon interactions including the Migdal effect \cite{DarkSide:2022dhx} and on dark matter particle interactions with electron final states \cite{DarkSide-50:2022hin}.  These limits may be improved in the future by better constraining the LAr ionization response  and the stochastic model underlying the NR quenching. 

\begin{acknowledgements}
The DarkSide Collaboration offers its profound gratitude to the LNGS and its staff for their invaluable technical and logistical support. We also thank the Fermilab Particle Physics, Scientific, and Core Computing Divisions. Construction and operation of the DarkSide-50 detector was supported by the U.S. National Science Foundation (NSF) (Grants No. PHY-0919363, No. PHY-1004072, No. PHY-1004054, No. PHY-1242585, No. PHY-1314483, No. PHY-1314501, No. PHY-1314507, No. PHY-1352795, No. PHY-1622415, and associated collaborative grants No. PHY-1211308 and No. PHY-1455351), the Italian Istituto Nazionale di Fisica Nucleare, the U.S. Department of Energy (Contracts No. DE-FG02-91ER40671, No. DEAC02-07CH11359, and No. DE-AC05-76RL01830), the Polish NCN (Grant No. UMO-2019/33/B/ST2/02884) and the Polish Ministry for Education and Science (Grant No. 6811/IA/SP/2018). We also acknowledge financial support from the French Institut National de Physique Nucl\'eaire et de Physique des Particules (IN2P3),   the  IN2P3-COPIN consortium (Grant No. 20-152),  and the UnivEarthS LabEx program (Grants No. ANR-10-LABX-0023 and No. ANR-18-IDEX-0001),  from the São Paulo Research Foundation (FAPESP) (Grant No. 2016/09084-0),  from the Interdisciplinary Scientific and Educational School of Moscow University ``Fundamental and Applied Space Research'',  from the Program of the Ministry of Education and Science of the  Russian  Federation  for  higher  education  establishments,  project No. FZWG-2020-0032 (2019-1569), from IRAP AstroCeNT funded by FNP from ERDF, and from the Science and Technology Facilities Council, United Kingdom.  I.~Albuquerque is partially supported by the Brazilian Research Council (CNPq). The theoretical calculation of beta decays was performed as part of the EMPIR Project 20FUN04 PrimA-LTD. This project has received funding from the EMPIR program co-financed by the Participating States and from the European Union’s Horizon 2020 research and innovation program. Furthermore, this project has received funding from the European Union’s Horizon 2020 research and innovation program under grant agreement No 952480. Isotopes used in this research were supplied by the United States Department of Energy Office of Science by the Isotope Program in the Office of Nuclear Physics.

 %\subsection*{\textit{Note added}} The analysis described in this work  was also applied  to  improve existing limits on spin-independent WIMP-nucleon interactions including the Migdal effect \cite{DarkSide:2022dhx} and on dark matter particle interactions with electron final states \cite{DarkSide-50:2022hin}. 

 \end{acknowledgements}

\bibliography{biblio}
\end{document}